\documentclass[aps,pra,twocolumn,showpacs,superscriptaddress,groupedaddress]{revtex4}

\usepackage{graphicx}
\usepackage{dcolumn}
\usepackage{bm}

\usepackage{amsmath} 
\usepackage{esint}
\usepackage{amsfonts}
\usepackage{mathrsfs} 
\usepackage{tabularx}  
\usepackage{booktabs}  
\usepackage{multirow}

\usepackage{braket}
\usepackage{color}
\usepackage[colorlinks,
linkcolor=blue,
anchorcolor=blue,
citecolor=blue]{hyperref}

\begin{document}

\title{Threats and Opportunities: Blockchain meets Quantum Computation}
\author{Wei Cui}
\email{aucuiwei@scut.edu.cn}
\address{School of Automation Science and Engineering, South China University of Technology, Guangzhou 510640, China}

\author{Tong Dou}
\address{School of Automation Science and Engineering, South China University of Technology, Guangzhou 510640, China}

\author{Shilu Yan}
\address{School of Automation Science and Engineering, South China University of Technology, Guangzhou 510640, China}

\affiliation{}


\begin{abstract}
	
This article considered deficiencies of the flourishing blockchain technology manifested by the development of quantum computation.  We show that the future blockchain technology would under constant threats from the following aspects: 1) Speed up the generation of nonces; 2) Faster searching for hash collisions; 3) Break the security of the classical encryption. We also demonstrate that incorporating some quantum properties into blockchain makes it  more robust and more efficient. For example people can establish a quantum-security blockchain system that utilizes quantum key distribution (QKD), and quantum synchronization and detectable Byzantine agreement (DBA)  can help  the blockchain systems achieve faster consensus even if there  exist a number of malicious nodes.

\end{abstract}

\maketitle

\section{Introduction}

 A blockchain is a distributed, transparent and timestamp ledger of cryptographically linked units of data. It consists of a sequence of blocks that are stored on and copied between publicly accessible servers. Each block consists of four fundamental elements: 1) the hash of the preceding block; 2) the data content of the block; 3) the nonce that is used to give a particular form to the hash; 4) the hash of the block. Generally speaking, a blockchain is a chain structure
that each block points to the previous one by a hash pointer and the transactions contained within the block are organized into a Merkel tree (Fig.1) to provide an efficient proof of existence.
Blockchain uses digital signatures and hash functions to verify data and identity, and redundant storage of multiple nodes to
ensure unmodifiable data \cite{Casino}.
 Since the 2008 pseudo-named
Nakamoto Satoshi released the Bitcoin white paper on the Internet \cite{Nakamoto}, various blockchain projects have emerged one after another. Although the characteristics of these projects are diverse and involve all aspects, in essence, the
most suitable scenario for blockchain is as a currency system and reputation system.  
 In this short article, we are only interested in the data structures and security algorithms related to blockchain. In particular, we consider the power of quantum computers and the capabilities of existing quantum algorithms \cite{Shor, Grover, Yao} represent a threat to the blockchain system. 
As it is pointed out by Fedorov, Kiktenko, and Lvovsky \cite{Fedorov}, blockchain technology as we know it today, may founder unless it integrates quantum technologies. While there are some interim solutions that incorporate post-quantum  cryptography, they do not guarantee unconditionally secure solutions to the threat.
 
 \begin{figure*}[!htb]
  \centering
  \includegraphics[width=\hsize]{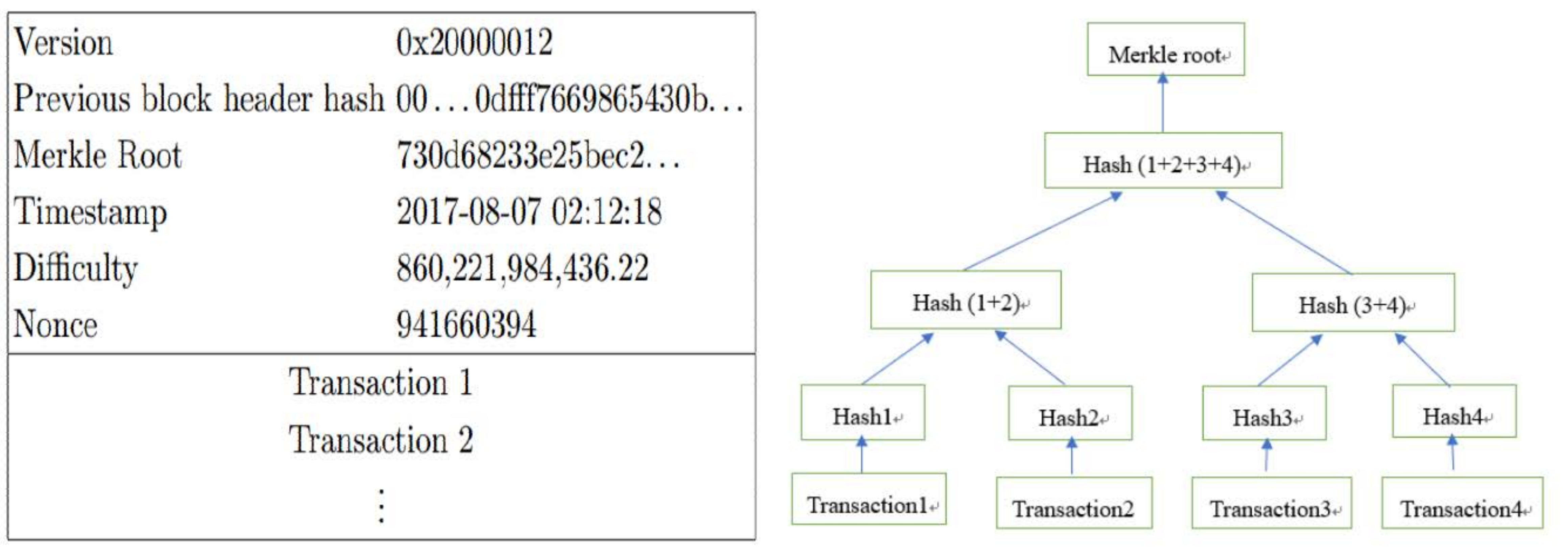}
  \caption{(Left) Illustration of a block. The data in the top constitutes the block header; (Right) Illustration of Merkle root. }
  \label{fig1}
\end{figure*}

Quantum computation differs from classical computation in two fundamental ways. First, quantum computing is not built on bits that are either zero or one, but on qubits that can be overlays of zeros and ones. Second, qubits not only exist in isolation but also become entangled and act as a group. It is quantum phenomena such as entanglement and superposition that enable quantum computers to perform operations that have no counterpart in the classical world. These are powerful mechanisms that give quantum computers a large advantage over classical computers. While the concept of quantum computer has been proposed for over 40 years, since Feynman first proposed them in 1980s, their most disruptive potential is still only conjecture, because building a universal quantum computer remains extremely difficult in engineering terms. However, in the past ten years, research underway at multiple big technology companies and startups, among IBM, Google, Rigetti, IonQ, Microsoft, Honeywell and D-wave, has led to a series of technological breakthroughs in building quantum computer systems. For example,  Google's Sycamore processor with 53 qubits takes about 200 seconds to sample one instance of a quantum circuit a million times \cite{Arute}. 
 Rigetti announced that it plans to build and deploy a 128-qubit system. Ion Q developed new trapped ion quantum computers with 160 stored and 79 processing qubits. Recently, Honeywell also reported on the integration of all necessary ingredients of the trapped-ion quantum charge-coupled device  architecture into a robust, fully-connected and programmable trapped-ion quantum computer \cite{Pino}. 
  It seems that we are getting closer to ``Quantum Supremacy". In the long term, quantum computers will very likely shape new computing and business paradigms by solving computational problems that are currently out of reach. Below we schematic introduce some quantum threats and opportunities to blockchain technology. 

\section{The threats posed by quantum computation}
\subsection{Speed up the generation of nonces and hash collisions}
Blockchain relies on the computation of hashes to provide security against modification of the past blocks as finding a hash collision with the existing hash requires a great cost. However, Grover's algorithm \cite{Grover}, a solution to the problem of finding a pre-image of a value of SHA-256 function that is difficult to invert, would give a speedup of $\sqrt{N}$ compared to classical collision search algorithm, so it would be much easier to find a hash collision than brute force searching. In Fig.2 and Fig.3 we show the Grover's algorithm and the Oracle of PoW model, respectively. 
 This potentially makes it viable to insert a modified block into the chain without compromising the sequential consistency of the blocks. On the other hand, since the longest chain is conventionally chosen as the accepted truth, the faster growing chain will come to dominate the blockchain and thus the faster miners can take control of the content of the blockchain, which is what we're talking about 51\% attack. With the speed increase provided by Grover's algorithm in the calculation of the nonce, it is feasible for a party with a quantum computer to rapidly outstrip competitors, who have only classical computing capacity, in generating additional blocks on the chain.

\subsection{Break the security of the classical encryption}
One of the first applications of quantum computation is to crack public-key cryptography and break digital signatures which are based on integer factoring problem and discrete logarithm problem. It means that the blockchain technology is facing threats from quantum computing. Take bitcoin for example, it uses elliptic curve digital signature algorithm which based on finding numbers on elliptic curves to ensure that funds can only be spent by their rightful owners. But with a modified version of the Shor algorithm \cite{Shor}, it is possible to determine all ECC-related keys used by bitcoin. Recently, a Google's research team found that it is possible to break RSA-2048 in 8 hours using 20 million noisy qubits \cite{Gidney}. Though their study is not aim at the arithmetic in elliptic curve groups, now is the time to focus on how to guarantee security of blockchain in the context of quantum computing.
 \begin{figure}[!htb]
  \centering
  \includegraphics[width=\hsize]{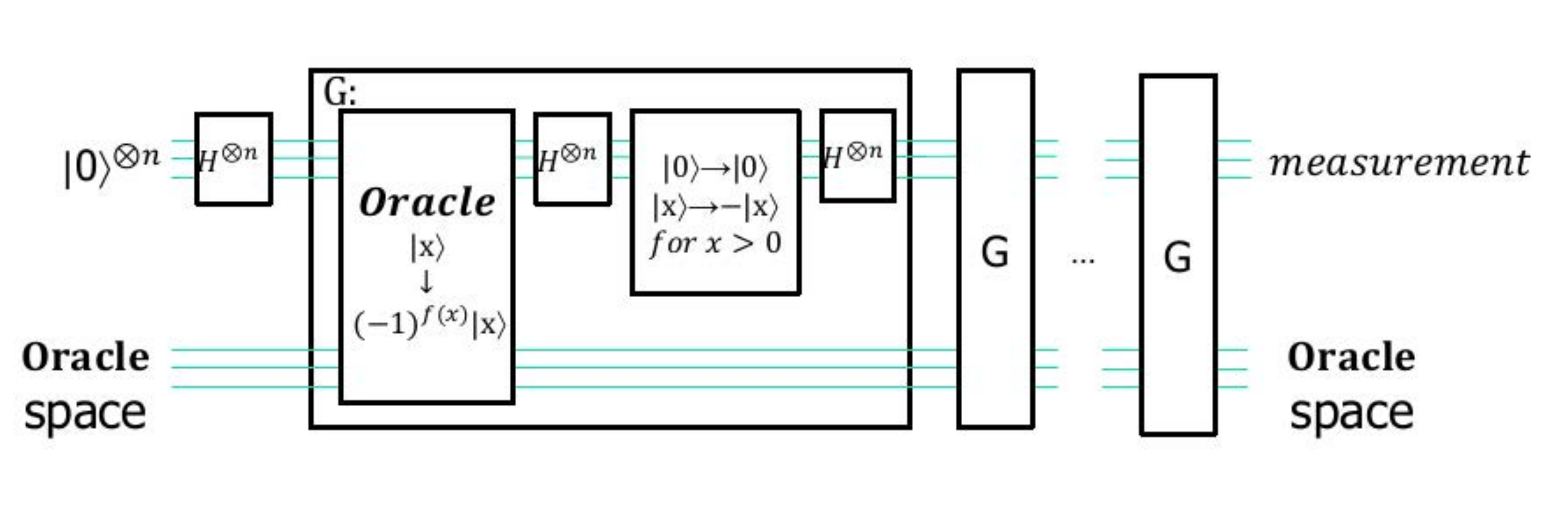}
  \caption{Grover's algorithm. }
\end{figure}

 \begin{figure}[!htb]
  \centering
  \includegraphics[width=\hsize]{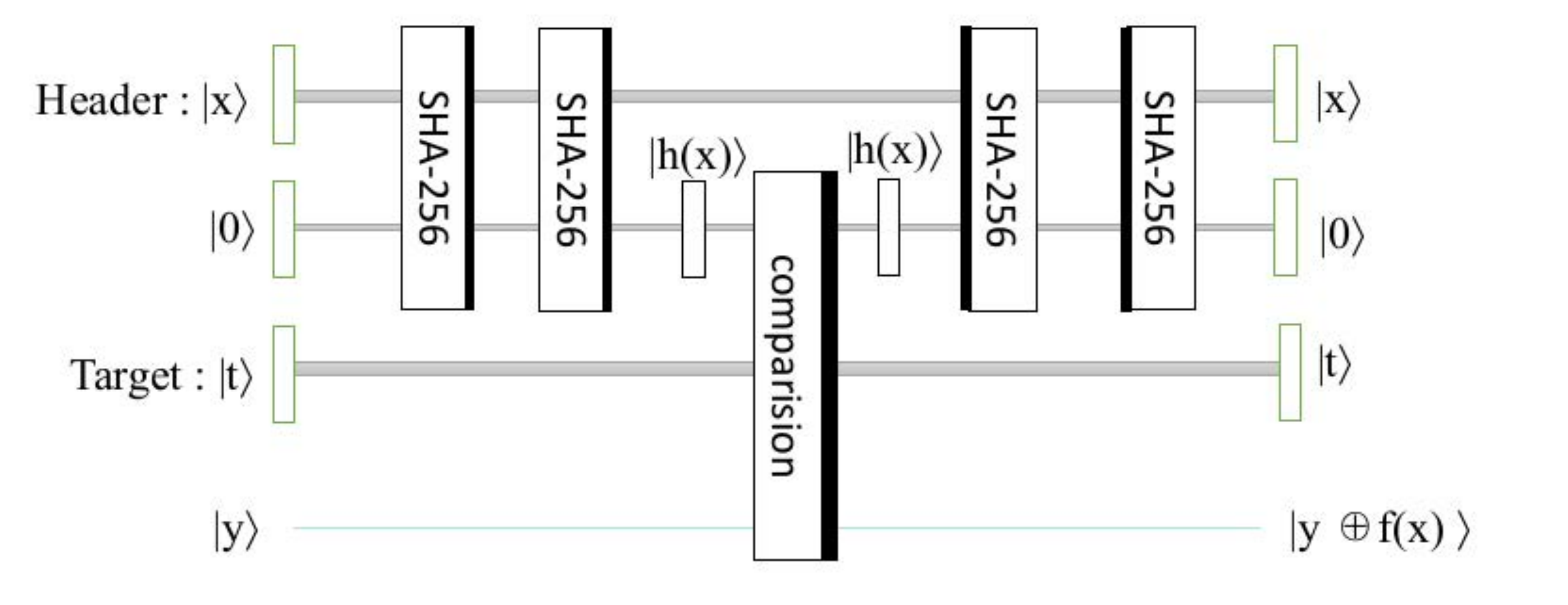}
  \caption{Oracle of PoW model.}
\end{figure}

\section{Opportunities presented by quantum technology}
\subsection{Quantum based cryptography--QKD}
 Not only is quantum computation bringing in a wave of innovation in technology and algorithm, but new ways of encryption are also being explored. There is a wide variety encryption methods that may make a significant impact to blockchain. Quantum key distribution (QKD) \cite{Preskill} is the primary and most mature technology in quantum cryptography which quantum computers cannot break. QKD is not based on mathematical complexity like normal methods of distributing cryptographic keys, but instead is secured by the laws of quantum mechanics: the process of measuring a quantum system in general disturbs the system. Though QKD do not transmit any message data, it can be used with any encryption algorithm to encrypt a message, and the encrypted message is considered provably secure. In addition to QKD, a novel method was proposed by Rajan and Visser, which can be considered as a blockchain of quantum version \cite{Rajan}. These technologies are still under development, and maybe become useful enough to protect blockchain against the attack from quantum computers in the future.

\subsection{Detectable Byzantine agreement and quantum synchronization}
 One of the most important concepts of blockchain is that it's a distributed system based on P2P networks, so it's necessary for it to solve the Byzantine agreement problem which pointed out the entire system couldn't reach a consensus when there exsits more than one-third faulty nodes. Many consensus algorithms used in blockchain are based on different application scenarios to select different algorithms to achieve consensus. For example, depending on the assumption that most nodes are honest, the bitcoin system uses the proof of work (PoW) algorithm to solve the Byzantine agreement problem probabilistically. Although the Byzantine agreement problem can't be solved absolutely in classical way, we can simplify it to the problem of generating and safely distributing correlative lists, and it finally evolves into detectable byzantine agreement (DBA). Quantum synchronization \cite{Rodenburg} can break through the limit to achieve consensus even if there are any number of faulty nodes in the system. Quantum solution to the detectable byzantine agreement could rely on the three qutrit entangled Aharonov state \cite{Fitzi} , or a QKD protocol \cite{Iblisdir}, or the four-qubit singlet state \cite{Gaertner}, or the single-qudit protocol \cite{Tavakoli}. 
The blockchain consensus algorithm is to ensure that the block information remains consistent. 
As we know the information that needs to be agreed on in the blockchain can be decomposed into binary strings up to arbitrary accuracy agreed upon in advance. So DBA could help to make each node in blockchain achieve a barrier-free consensus of the system through the correlative list. As an example, one can use the GHZ state 
$$GHZ=\frac{{\vert0\rangle}^{\otimes N}+{\vert1\rangle}^{\otimes N}}{\sqrt{2}}$$
to achieve distributed consensus with  the following steps:

(i) Prepare the GHZ state by entangling a set of $N$ qubits;

(ii) For $N$ processes, distribute a single one of the $N$ qubits from the superposition to
each one of the participants in the system;

(iii) Each participant measures their individual qubit;

(iv) Chose $0$ if the qubit measurement is $\vert0\rangle$ and chose $1$ if the qubit measurement is $\vert1\rangle$.

\section{Conclusion}

In summary, this article introduces deficiencies of the blockchain technology manifested by the development of quantum computation. We also show that quantum key distribution and quantum synchronization can help blockchain more security and more efficient. Although quantum computing is now still at the preliminary stage, it holds out the promise of computers with  tremendous power. 
We believe that both blockchain and quantum computation are subjects of active research and may show interesting development in the  next decade.

\begin{acknowledgments}
This work was supported by the National Natural Science Foundation of China under Grant 61873317, Guangdong Basic and Applied Basic Research Foundation under Grant 2020A1515011375,  and in part by Fundamental Research Funds for the Central Universities.

\end{acknowledgments}

\nocite{*}


\begin{thebibliography}{0}
\bibitem{Casino}
F. Casino, T. K. Dasaklis, and C. Patsakis,  A systematic literature review of blockchain-based applications: Current status, classification
and open issues, \emph{ Telematics and Informatics},  36: 55–81, 2019.



\bibitem{Nakamoto}
S. Nakamoto, Bitcoin: a peer-to-peer electronic cash system, https://bitcoin.org/bitcoin.pdf

\bibitem{Shor}
P. W. Shor, Polynomial-time algorithms for prime factorization and discrete logarithms on a quantum computer, \emph{SIAM J. Comput.},   26, 1484–1509, 1997.

\bibitem{Grover} 
 L. K. Grover, Quantum mechanics helps in searching for a needle in a Haystack, \emph{Phys. Rev. Lett.}, 79(2): 325-328, 1997. 
 
\bibitem{Yao}
J. Yao, M. Bukov, L. Lin,  Policy gradient based quantum approximate optimization algorithm, arXiv: 2002.01068, 2020. 

 
 \bibitem{Fedorov}
A. K. Fedorov,  E. O. Kiktenko, A. I. Lvovsky, Quantum computers put blockchain security at risk, \emph{Nature}, 563: 465–467, 2018.

\bibitem{Arute}
F. Arute, K. Arya, R. Babbush, et al., Quantum supremacy using a programmable superconducting processor, \emph{Nature}, 574: 505-511, 2019.
 
 \bibitem{Pino}
 J. M. Pino, J. M. Greiling, C. Figgatt, et al., Demonstration of the QCCD trapped-ion quantum computer architecture, https://www.honeywell.com/en-us/newsroom/news/2020/03/scientific-paper-demonstration-of-the-qccd-trapped-ion-quantum , 2020. 
 
\bibitem{Gidney} 
 C. Gidney, M. Eker\aa , How to factor 2048 bit RSA integers in 8 hours using 20 million noisy qubits, arXiv:1905.09749, 2019.
 
\bibitem{Preskill}
P. W. Shor and J. Preskill, Simple proof of security of the BB84 quantum key distribution protocol, \emph{Phys. Rev. Lett.}, 85: 441-444, 2000. 
 
 
 
\bibitem{Rajan} 
 D. Rajan, M. Visser, Quantum blockchain using entanglement in time, \emph{Quantum Report}, 1(1):3-11, 2019.
 
 \bibitem{Rodenburg}
B. Rodenburg and S. P. Pappas,  Blockchain and quantum computing, NJ Princeton, https://www.mitre.org/publications/technical-papers/blockchain-and-quantum-computing .


 \bibitem{Fitzi}
M. Fitzi, N. Gisin, U. Maurer, Quantum solution to the Byzantine agreement problem, \emph{Phys. Rev. Lett.},  87(21):217901, 2001.
 
 \bibitem{Iblisdir}
S. Iblisdir, N. Gisin,  Byzantine agreement with two quantum-key-distribution setups, \emph{Phys. Rev. A},   70(3):034306, 2004.

\bibitem{Gaertner}
S. Gaertner, M. Bourennane, C. Kurtsiefer, et al., Experimental demonstration of a quantum protocol for Byzantine agreement and liar detection, \emph{Phys. Rev. Lett.}, 100(7):070504, 2008.

\bibitem{Tavakoli}
A. Tavakoli, A. Cabello, Z. Marek, et al., Quantum clock synchronization with a single qudit, \emph{Scientific Reports},  5:7982, 2015.
	
	
\end{thebibliography}

\end{document}